\RequirePackage{fix-cm}
\documentclass[twocolumn]{svjour3}       
\smartqed  
\usepackage{enumitem}
\usepackage{url}
\usepackage{algorithm}
\usepackage{algorithmicx}
\usepackage{algpseudocode}
\usepackage{latexsym}
\usepackage{color}
\usepackage{epsfig}
\usepackage{subfigure}
\usepackage{amsfonts}
\usepackage{amssymb}
\newcommand{\etal}[0]{{\it et al. }}

\usepackage{epstopdf}
\AppendGraphicsExtensions{.tif}
\usepackage{mathtools,etoolbox}
\begin{document}

\title{Functional Segmentation through Dynamic Mode Decomposition: Automatic Quantification of Kidney Function in DCE-MRI Images\thanks{Santosh Tirunagari and Norman Poh have benefited  from  the  Medical  Research  Council  (MRC) funded project ``Modelling the Progression of Chronic Kidney Disease'' under the grant number R/M023281/1.}
}

\titlerunning{Movement Correction in DCE-MRI via WR-DMD}        

\author{Santosh~Tirunagari \and Norman~Poh \and Kevin~Wells \and Miroslaw~Bober \and Isky~Gorden \and David~Windridge}

\authorrunning{Tirunagari \etal} 


\institute{S. Tirunagari \and N. Poh
\at Department of Computer Science, University of Surrey, Guildford, Surrey GU2 7XH, UK.\\Tel.: +44-(0)1483-6179.
\and
S. Tirunagari \and K. Wells \and M. Bober
\at Center for Vision, Speech and Signal Processing (CVSSP), University of Surrey, Guildford, Surrey GU2 7XH, UK.
\and
I. Gorden \at University College London (UCL) Institute of Child Health, 30 Guildford Street, London WCIN lEH, UK.
\and
D. Windridge \at Department of Computer Science, Middlesex University, The Burroughs, Hendon, London NW4 4BT, UK.\\
\\\email{\{s.tirunagari, n.poh, k.wells, m.bober\}@surrey.ac.uk}   
}

\date{Received: date / Accepted: date}

\maketitle

\begin{abstract}
Quantification of kidney function in Dynamic Contrast Enhanced Magnetic Resonance imaging (DCE-MRI) requires careful segmentation of the renal region of interest (ROI). Traditionally, human experts are required to manually delineate the kidney ROI across multiple images in the dynamic sequence. This approach is costly, time-consuming and labour intensive, and therefore acts to limit patient throughout and acts as one of the factors limiting the wider adoption of DCR-MRI in clinical practice.

Therefore, to address this issue, we present the first use of Dynamic Mode Decomposition (DMD) as a the basis for automatic segmentation of a dynamic sequence, in this case, kidney ROIs in DCE-MRI. Using DMD coupled combined with thresholding and connected component analysis is first validated on synthetically generated data with known ground-truth, and then applied to ten healthy volunteers’ DCE-MRI datasets.

We find that the segmentation result obtained from our proposed DMD framework is comparable to that of expert observers and very significantly better than that of an a-priori bounding box segmentation. Our result gives a mean Jaccard coefficient of 0.87, compared to mean scores of 0.85, 0.88 and 0.87 produced from three independent manual annotations. This represents the first use of DMD as a robust automatic data driven segmentation approach without requiring any human intervention. This is a viable, efficient alternative approach to current manual methods of isolation of kidney function in DCE-MRI.

\end{abstract}


\section{Introduction}
Diagnosis of renal dysfunction based on blood and urine tests often produces inaccurate results as the creatinine levels in blood are detectable only after 60\% of the renal dysfunction has taken place~\cite{zollner2007assessment}. Therefore, to address this limitation Dynamic Contrast Enhanced Magnetic Resonance Imaging (DCE-MRI) has been proposed~\cite{tofts2010t1,sourbron2013classic}. DCE-MRI is a non-ionising alternative to conventional radioisotope renography. It has particular attraction in cases of Chronic Kidney Disease (CKD), which requires repeated functional assessment, as well as in paediatric cases where exposure to repeated radiation doses is of greater concern than in adults~\cite{khrichenko2010functional,michaely2007functional,prasad2006functional}. A further benefit of DCE-MRI is that anatomical images can also be obtained during the same imaging session, providing a direct comparisons to the observed physiological abnormalities.
 
\begin{figure}[h]
\centering
\begin{tabular}{c}
\includegraphics[scale=0.5]{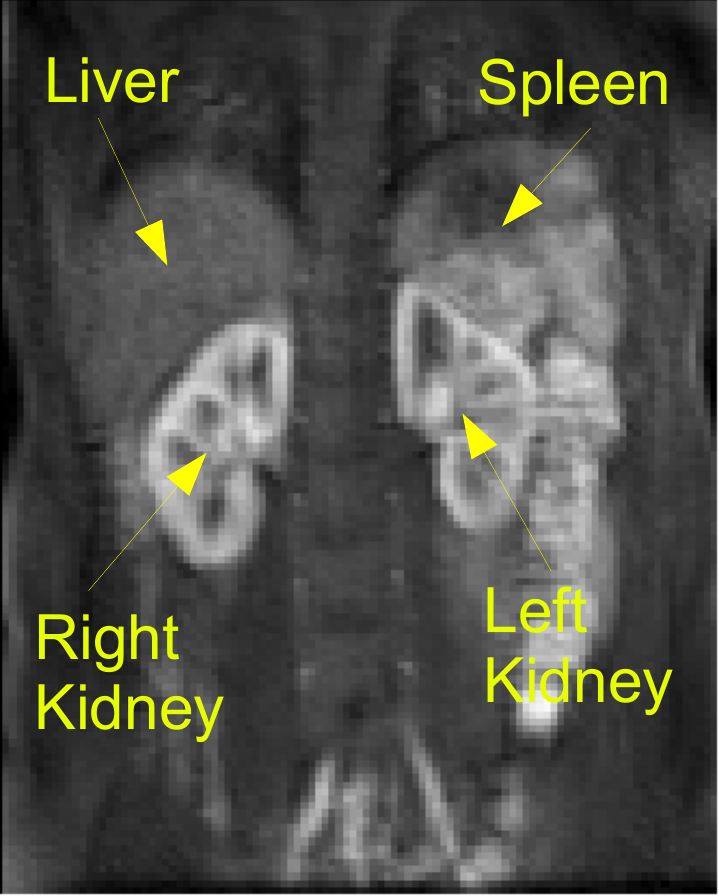}\\
\end{tabular}
\caption{DCE-MRI image with anatomical parts: \emph{kidney}, \emph{liver} and \emph{spleen}. Renal perfusion takes place inside the kidney regions after the injection the contrast agent. The kidney function is then quantified by calculating the mean pixel intensity values inside the kidney region (in practice using a tracer-kinetic method~\cite{tofts2012precise}). For this purpose, a proper segmentation of kidney region is required.}
\label{anatomy}
\end{figure}

Absolute quantification of kidney (renal) function in DCE-MRI (see Figure~\ref{anatomy}) is often obfuscated as it requires manual segmentation~\cite{zollner2009assessment,zollner2012assessment} of the kidney ROI, (i.e., a region of kidney is selected as a template by human experts by manually delineating the kidney ROI). Semi-automatic methods, however, work by specifying the target ROI a-\emph{priori} to an automated segmentation algorithm~\cite{huang2004functional}. Although these approaches are potentially correct, the major issue is the need for human intervention in the segmentation process of the target region. In addition, this can be labour intensive, time-consuming and inefficient as the human expert has to examine the whole sequence of images to find the most suitable frame. This is typically also inconvenient since human experts require proprietary software for delineating the ROI, and also error-prone as the selection of the template (ROI) is subjected to observer variations~\cite{de2000mr}.
 
Automated methods~\cite{johnson2011determinants,rusinek2007performance} have the potential to overcome these limitations and  moreover offer a more reproducible approach. In order to achieve a complete automatic approach reliably, we propose to use Dynamic Mode Decomposition (DMD), which has been used extensively in modelling fluid dynamics. Despite the complexity of the dynamics of an image sequence containing anatomical structures, the information dynamics in our approach are represented in an extremely efficient manner within individual ``modes''. These experiments show, for the first time, that DMD can capture distinctive features that clearly distinguish various functional segments within a dynamic MRI image sequence. In this paper, we thus report in detail a framework utilising DMD in conjunction with a simple thresholding technique to obtain functional segmentation of the kidney ROI.  

Previously, automated segmentation methods utilising clustering and classification methods such as k-means clustering~\cite{zollner2009assessment} and k-nearest neighbour classification~\cite{hodneland2011vivo} have been suggested. These methods work by considering the signal intensity values across the images in time, thus obtaining a high-dimensional feature vector in each voxel based on the actual tissue response before and after injecting the contrast agent. Methods based on active contours~\cite{abdelmunim2008kidney} and related methods have also offered solutions that consider the region boundary coupled with shape constraints~\cite{ali2007graph}. A recent work by Hodneland \etal in 2014 applied the temporal tissue response and minimal boundary length as shape information for obtaining the kidney segmentation~\cite{hodneland2014segmentation} in 4D DCE-MRI videos. 
  
Zollner et al~\cite{zollner2007assessment} introduced the Independent Component Analysis (ICA) technique for functional segmentation of human kidney ROI in DCE-MRI recordings. 
An approach based on spatio-temporal ICA (STICA)~\cite{kiani2012line} was also developed recently that offers a fully data-driven approach exploiting the distribution of the properties of the spatial data incrementally in the direction of the time axis. A major limitation for this ICA based approach is finding an optimal filter that maximises the statistical independence of the observed signals. The ICA method is typically also approached with a substantial number of assumptions/heuristics and is computationally expensive. 

In our previous studies~\cite{Tirunagari2017}, we presented a novel automated, registration-free movement correction approach
based on windowed and reconstruction variants of Dynamic Mode Decomposition (WR-DMD) to suppress unwanted complex organ motion in DCE-MRI image sequences caused due to respiration.

Our methodological framework in~\cite{Tirunagari2017} consisted of the following steps:1) DCE-MRI sequence consisting of $N$
images was processed using the windowed DMD (W-DMD) algorithm in order to output each $N − 2$ W-DMD components C1 and C2. At this stage the W-DMD(C1) produced the low rank images and W-DMD(C2) produced sparse images. 2) W-DMD(C1) was then given as an input to DMD which produced $N − 3$ DMD modes. The first $3$ DMD modes were then selected for reconstructing the motion stabilised image sequence. Out of the first three dynamic modes, the first mode revealed a low-rank model and
the remaining $N-4$ modes captured the sparse representations. The contrast changes were captured in the most significant modes, in particular, mode-2 was captured kidney region and mode-3 and 4, captured spleen and the liver regions respectively. Noise and residuals including the motion components were captured in the remaining of the modes. Therefore, this study investigates whether we could perform segmentation of the kidney region of interest from dynamic mode-2 for automatic quantification the kidney function.

Dynamic mode decomposition, due to its ability to identify regions of dominant motion in an image sequence in a completely data-driven manner without relying on any prior assumptions about the patterns of behaviour within the data, has gained significant applications in various fields~\cite{tirunagari2015windowed,tirunagari2016can,grosek2014dynamic,6926317,2015arXiv150804487M,brunton2016extracting,tirunagari2015detection}.
Therefore, it is thus potentially be well-suited to analyse a wide variation of blood flow and filtration patterns seen in renography pathology~\cite{Tirunagari2017}. 

The novelty of our proposal thus lies in utilising DMD to carry out functional segmentation from medical image sequences in a manner that is both extremely efficient and completely data driven (and thus heuristic-free). Extracting relevant key organs from the DCE-MRI images can potentially provide clinicians with a better way to manage clinical time and cost. 

The remainder of this paper is organised as follows: Section~\ref{method} considers the theory for DMD as well as presenting  evaluation criteria based on Jaccard Similarity Coefficient. Section~\ref{data} describes datasets that are based on synthetic data and a set of 10 healthy volunteers' MRI datasets. Section~\ref{exp_res} derives our experimental objectives and presents the results and, finally, conclusions are drawn and are discussed in Section~\ref{conc}.  

\section{Methodology}
\label{method}
 
\begin{figure*}
\centering
\includegraphics[scale=0.6]{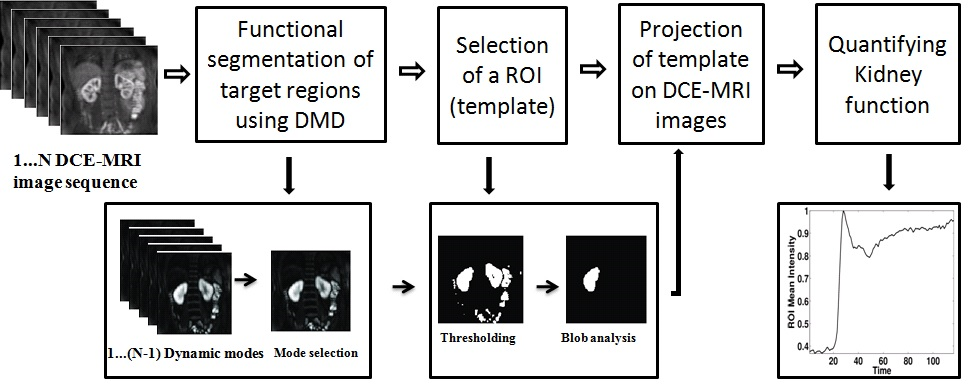}
\caption{Flow chart showing the steps involved in the methodological framework. First, a DCE-MRI
sequence consisting of $N$ images is processed using the DMD
algorithm in order to output $N-1$ dynamic mode images.
From these, we select a dynamic image corresponding to
mode 2. Secondly, thresholding is performed for the obtained dynamic mode image and converted to binary image. After
binarization, the largest area with connected pixels is selected
as a template. Finally, the produced template is projected onto
DCE-MRI data to compute the kidney function.}
\label{flowchart}
\end{figure*}

In this section, we present our methodological framework which consecutively consists of DMD, thresholding, selection of kidney regions using connected component analysis and finally modelling of the kidney function. The overall process pipeline is shown in Figure~\ref{flowchart}. At the end of this section, we also present our evaluation criteria i.e., Jaccard similarity coefficient for comparing segmentation results.

\subsection{Dynamic Mode Decomposition (DMD)}
\label{meth:DMD}

Let $\bar{\mathbf{x}}_r$ be the $r^{th}$ dynamic image frame in a DCE-MRI sequence, whose size is $m\times n$. This image frame $\bar{\mathbf{x}}_r$ is converted to $mn\times1$ column vector, resulting in the construction of a data matrix $\textbf{X}$ of size  $mn \times N$ for $N$ image frames.

\begin{equation}\label{eq:display13} 
\textbf{X} =   [\bar{\mathbf{x}}_1, \bar{\mathbf{x}}_2, \bar{\mathbf{x}}_3, \cdots, \bar{\mathbf{x}}_N] = \begin{pmatrix}
x^{1}_1 & x^{1}_2 & ... & x^{1}_N \\
\vdots & \vdots  &  \vdots & \vdots  \\  
x^{mn}_1 & x^{mn}_2 & ... & x^{mn}_N \\
\end{pmatrix}.
\end{equation}

The images in the DCE-MRI data are collected over regularly spaced time intervals and hence each pair of consecutive images are correlated. It can be justified that a mapping $A$ exists between them forming a span of krylov subspace~\cite{krylov1931numerical,saad1981krylov,ruhe1984rational}: 

\begin{equation}
\label{ch_2_A_Label}
 \begin{split}
    \textbf{X} = [\bar{\mathbf{x}}_1, A\bar{\mathbf{x}}_1, A^2\bar{\mathbf{x}}_1, A^3\bar{\mathbf{x}}_1, \cdots, A^{N-1}\bar{\mathbf{x}}_{1}].\\
    [\bar{\mathbf{x}}_2, \bar{\mathbf{x}}_3, \cdots, \bar{\mathbf{x}}_N] = A[\bar{\mathbf{x}}_1, \bar{\mathbf{x}}_2, \cdots, \bar{\mathbf{x}}_{N-1}]. \\
    \mathbf{P}_2 = A\mathbf{P}_1 + \mathbf{re}^{T}_{N-1}.
  \end{split}
\end{equation}

Here, ${\mathbf r}$ is the vector of residuals that accounts for behaviours that cannot be described completely by ${\mathbf A}$, $\mathbf{e}_{N-1} = \{0, 0 , … , 1\} \in  \mathcal{R}^{N-1}$, $ \mathbf{P}_2 =  [\bar{\mathbf{x}}_2, \bar{\mathbf{x}}_3, \cdots, \bar{\mathbf{x}}_N] $ and   $ \mathbf{P}_1 = [\bar{\mathbf{x}}_1, \bar{\mathbf{x}}_2, \cdots, \bar{\mathbf{x}}_{N-1}]$. The system $\mathbf{A}$ is unknown and it captures the overall dynamics within the dynamic image sequence in terms of the eigenvalues and eigenvectors of $\mathbf{A}$, which are referred to as the DMD eigenvalues and DMD modes respectively. 

The sizes of the matrices $P_2$ and $P_1$ are both $mn \times N-1$ each. Therefore, the size of unknown matrix $A$ would be $mn \times mn$.
Since, matrix $A$ captures the dynamical information within the image sequence, solving it provides us with the dynamics captured in the image sequence in terms of modes. Unfortunately, solving for A is computationally very expensive due to it size. For instance, if an image has a size of $240 \times 320$ i.e., $m=240$ and $n=320$, the size of $A$ is then $76800 \times 76800$. 

From the literature there are two approaches for obtaining these eigenvalues and modes. The first is Arnoldi based approach, which is useful for theoretical analysis due to its connection with Krylov subspace methods~\cite{krylov1931numerical,saad1981krylov,ruhe1984rational}. The second is a singular value decomposition (SVD) based approach that is more robust to noise in the data and to numerical errors~\cite{Schmid3}.

Therefore, SVD can be directly applied on $\mathbf{\mathbf{P}_1}$ in Eq.~(\ref{ch_2_A_Label}), to obtain $\mathbf{U}$, $\mathbf{\Sigma}$ and $\mathbf{V^*}$ matrices.

\begin{equation}\label{eq:ar3}
\mathbf{\mathbf{P}_2} = \mathbf{A U \Sigma V^*}.
\end{equation}
$\because \mathbf{\mathbf{P}_1}=\sum\limits_{i=1}^d\Sigma_i\mathbf{u}_i\mathbf{v}_i^{*}$,

here, $\Sigma_{i}$ is the $i^{th}$ singular value of $\mathbf{P}_1$, $\mathbf{u}_{i}$ and $\mathbf{v}_{i}^*$ are the corresponding left and right singular vectors respectively; and $d$ is the total number of singular values.

Re-arranging Eq.~(\ref{eq:ar3}), we  obtain the the full-rank matrix $\mathbf{{A}}$,   

\begin{equation}\label{eq:ar4}
\mathbf{A} = \mathbf{\mathbf{P}_2V\Sigma^{-1}U^*}.
\end{equation}

Since the eigenvalue analysis is agnostic to any linear projection; so solving the eigen problem of $\mathbf{\widetilde{H}}$ is easier than that of solving for $\mathbf{A}$ directly. Moreover, the associated eigenvectors of $\mathbf{\widetilde{H}}$ provide the coefficients for the linear combination that is necessary to express the dynamics within the time series basis.
\begin{equation}
\label{sigmavalues}
\mathbf{\widetilde{H}}\mathbf{\omega} = \mathbf{\sigma}\mathbf{\omega},
\end{equation}
where, $\mathbf{\omega}$ are the eigenvectors and $\mathbf{\sigma}$ a diagonal matrix containing the corresponding eigenvalues of $\mathbf{\widetilde{H}}$ matrix. The eigenvalues of $\mathbf{\widetilde{H}}$ approximate some of the eigenvalues of the full system $\mathbf{A}$~\cite{DBLP:journals/corr/GrosekK14}, we then have:
\begin{equation}
\label{dmv_der}
\begin{aligned}
  \mathbf{AU} &\approx \mathbf{U\widetilde{H}},\\
  \mathbf{AU} &\approx \mathbf{U \omega \sigma \omega^{-1}},\\
  \mathbf{A(U \omega)} &\approx \mathbf{(U \omega)\sigma}.\\
\end{aligned}
\end{equation}
Therefore $\mathbf{\widetilde{H}}$ is determined  on  the  subspace  spanned by the  orthogonal singular basis  vectors $\mathbf{U}$ obtained via $\mathbf{\mathbf{P}_1}$,
\begin{equation}
\begin{aligned}
  \mathbf{\widetilde{H}} & = \mathbf{U^*(A)U},\\
  \mathbf{\widetilde{H}} & = \mathbf{U^*(\mathbf{P}_2V\Sigma^{-1}U^*)U},
\end{aligned}
\end{equation}
which can be rewritten as: 
\begin{equation}
\mathbf{\widetilde{H}} = \mathbf{U^*\mathbf{P}_2V\Sigma^{-1}}.
\end{equation}
Here $\mathbf{U^*} \in \mathbb{C}^{(N-1) \times mn}$ and $\mathbf{V} \in \mathbb{C}^{(N-1) \times (N-1)}$ are the conjugate transpose of $\mathbf{U}$ and $\mathbf{V^*}$, respectively; and $\mathbf{\Sigma^{-1}} \in \mathbb{C}^{(N-1) \times (N-1)}$ 
denotes the inverse of the singular values $\mathbf{\Sigma}$. 
By replacing $\mathbf{\Psi} = \mathbf{U \omega}$ in Eq.~(\ref{dmv_der}) i.e., $\mathbf{A(\Psi)} \approx \mathbf{(\Psi)\sigma}$, we obtain the dynamic modes $\mathbf{\Psi}$. $\because \mathbf{U} = \mathbf{\mathbf{P}_2V\Sigma^{-1}}$, therefore, we have:
\begin{equation}
\label{dmv}
\mathbf{\Psi} = \mathbf{\mathbf{P}_2V\Sigma^{-1}\omega}
\end{equation}

The complex eigenvalues $\mathbf{\sigma}$ contain growth/decay rates and frequencies of the corresponding DMD modes \cite{Schmid2,Schmid1}. If $\sigma_j$ are the diagonal elements of $\mathbf{\sigma}$ from Eq.~(\ref{sigmavalues}), the temporal behaviour of the DMD modes is then formed via Vandermonde matrix $\mathcal{V}$, which raises its column vector to the appropriate power. $\mathcal{V}(f)$ with $(N-1)\times (f+1)$ elements will be defined by the following:
\begin{equation}
\mathcal{V}(f) = \begin{pmatrix}
1 & \sigma_1^1 & \sigma_1^2 & ... & \sigma_1^f   \\
1 & \sigma_2^1 & \sigma_2^2 & ... & \sigma_2^f  \\
\vdots & \vdots & \vdots &  \vdots & \vdots \\  
1 & \sigma_{N-1}^1 & \sigma_{N-1}^2 & ... & \sigma_{N-1}^f \\
\end{pmatrix},
\end{equation}
$\mathcal{V}(N)$ is a standard Vandermonde matrix for reconstruction but if $f>N$, this is used for forecasting.
DMD modes with frequencies $\mu_{j}$ is defined by:
\begin{equation}
\mu_{j} = \frac{\ln(\sigma_{j})}{\delta{t}},
\end{equation}
where $\delta{t}$ is the lag between the images. The real part of $\mu_{j}$ regulates the growth or decay of the DMD modes, while the imaginary part of $\mu_{j}$ drives oscillations in the DMD modes.

\subsection{Ordering Dynamic modes}
\label{rank}

In order to select the most significant dynamic modes, the method suggested in~\cite{kutz2016multiresolution,grosek2014dynamic} is to calculate the logarithmic values of the $diag(\sigma)$. The frequencies which are near origin are the most significant modes. 
The other way we propose here is by calculating the phase-angles for the complex eigenvalues. 

The absolute value for the phase-angles are calculated and modes with unique phase-angles are selected. Doing this will remove one of the conjugate pairs in the dynamic modes. These conjugate modes have same phase-angles but with different signs and look and capture similar information~\cite{sayadi2013dynamic}. After discarding one of the conjugate pairs, the dynamic modes are then sorted in ascending order of their phase-angles. The resultant dynamic modes are thus sorted according to their significance. In this study we have considered the first three significant dynamic modes when reconstructing the original sequence.

\begin{figure*}
\centering
\begin{tabular}{c}
\includegraphics[scale=1]{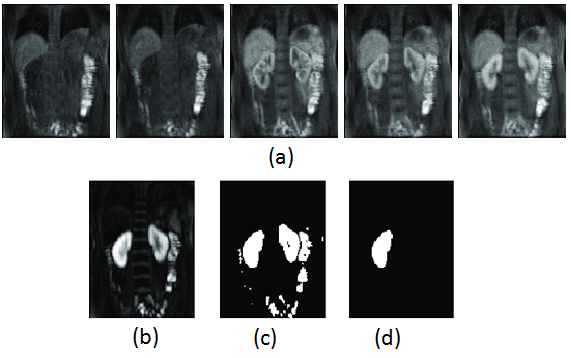}\\
\end{tabular}
\caption{(a) 5 randomly selected images from a DCE-MRI sequence of a healthy volunteer. The middle image shows the peak stage of renal perfusion inside the kidneys. Quantification of kidney function is performed by calculating the mean of pixel intensity values inside the kidneys. (b) Image showing their corresponding dynamic mode-2. (c) Image showing the binarized version of that dynamic mode. (d) Kidney template with the largest area of connected pixels.}
\label{dmd_pot}
\end{figure*}

The effectiveness of DMD, as a preliminary analysis, is demonstrated in Figure~\ref{dmd_pot}. The top five figures show random DCE-MRI images in temporal order for a healthy volunteer. Since renal perfusion is taking place inside the kidneys region, one can observe that the corresponding DMD mode in (b) is highlighting the dynamic changes inside the kidney region. A simple thresholding and binarization of the DMD mode thus highlights the kidney region as shown in (c). Selecting the region, or the largest  area of connected pixels, can thus give us the kidney template in an exemplar 4D dynamic medical imaging application.

\subsection{Performance Measure}
To evaluate the performance of our segmentation we use Jaccard similarity coefficient, a standard measure of similarity between finite sample sets~\cite{levandowsky1971distance} (here, in particular, the similarity between the two segmented sets). It is defined as the size of the intersection (of pixels values) between the segmented sets divided by the size of the union of the segmented sets: 

\begin{equation} \label{jaccard}
J(A,B) = \frac{|A\cap B|}{|A\cup B|}
\end{equation}

Here A and B are two segmented images. Values of the \emph{Jaccard index} range from 0 to 1. (0 if the intersection of the two sets is empty; 1 if the two sets are equal; the more similar the sets are, the closer to 1 is the metric).

\subsection{Evaluation Criteria}
\label{eval_crit}
Let $e_i$ and $d$ $\forall \ i \in \{1,2,3\}$ be the segmented results from the three human experts and DMD framework respectively. For an unbiased evaluation criterion, let us calculate three different ground-truths $G_i$ based on the segmentation results $e_i$ obtained from the human experts.
\begin{equation} 
\begin{split}
G_1 = (e_2\cap e_3),\\
G_2 = (e_1\cap e_3) \mbox{\ \&}\\
G_3 = (e_2\cap e_1).
\end{split}
\end{equation}

The ground-truth for the human expert-1 is the mutual agreement of segmentation results from expert-2 and expert-3. Similarly the ground-truth for expert-2 can be calculated as the mutual agreement of expert-1 and expert-3 and vice versa. The evaluation criteria $E_i$ for the human experts can be calculated as follows:

\begin{equation} \label{jaccard}
E_i = J(e_i,G_i) = \frac{|e_i\cap G_i|}{|e_i\cup G_i|} \ \forall \ i \in \{1,2,3\}.
\end{equation}

The evaluation $\mathbf{D}$ for the DMD framework is given by the average of $D_i  \ \forall \ i \in \{1,2,3\}$:
\begin{equation}
\mathbf{D} = \frac{1}{3}\sum_{i=1}^{3} D_i.
\end{equation}
where, 
\begin{equation}
D_i = J(d,G_i) = \frac{|d\cap G_i|}{|d\cup G_i|} \ \forall\  i \in \{1,2,3\}.
\end{equation}

\section{Dataset}
\label{data}
In this section we briefly describe the datasets that we have used in this study i.e., DCE-MRI data and synthetically generated data. 

\subsection{DCE-MRI data}
\begin{figure*}
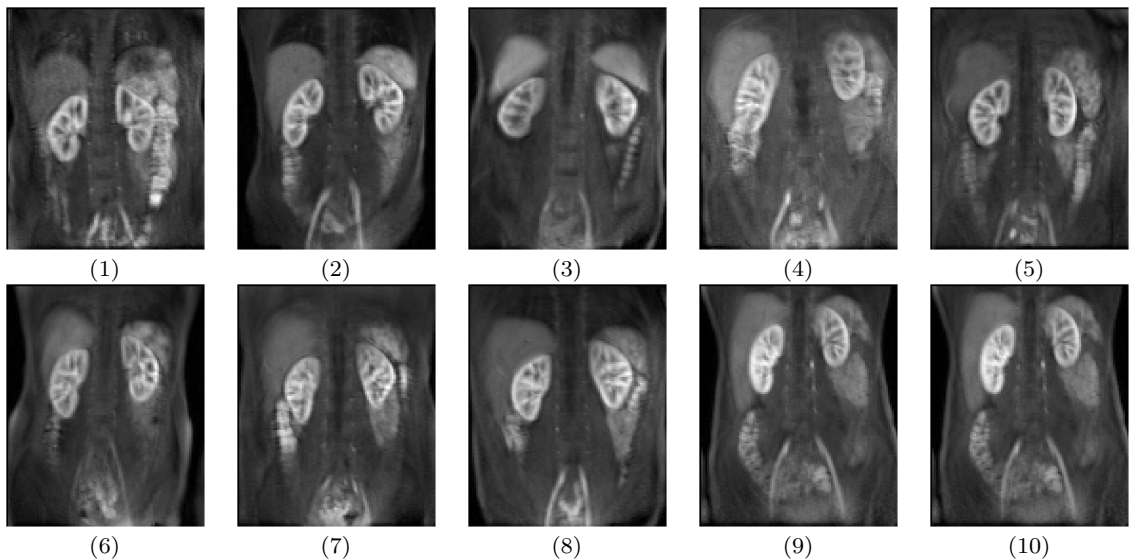

\centering
\begin{tabular}{ccccc}
\includegraphics[scale=0.35]{pictures/dataset_images/Volunteer-1.tif} & \includegraphics[scale=0.35]{pictures/dataset_images/Volunteer-2.tif} & \includegraphics[scale=0.35]{pictures/dataset_images/Volunteer-3.tif} & \includegraphics[scale=0.35]{pictures/dataset_images/Volunteer-4.tif} & \includegraphics[scale=0.35]{pictures/dataset_images/Volunteer-5.tif}\\
(1) & (2) & (3) & (4) &(5) \\
\includegraphics[scale=0.35]{pictures/dataset_images/Volunteer-6.tif} & \includegraphics[scale=0.35]{pictures/dataset_images/Volunteer-7.tif} & \includegraphics[scale=0.35]{pictures/dataset_images/Volunteer-8.tif} & \includegraphics[scale=0.35]{pictures/dataset_images/Volunteer-9.tif} & \includegraphics[scale=0.35]{pictures/dataset_images/Volunteer-10.tif}\\
(6) & (7) & (8) & (9) &(10) \\
\end{tabular}
\caption{Exemplars of Dynamic MR images from 10 healthy volunteers' kidney slice produced by DCE-MRI sequence considered as 10 different datasets in this study. The images here show the central kidney slice at time $120s$ aortic peak enhancement after the contrast agent is injected. The yellow boundary on the kidneys are a results of manual delineation from a human expert. The mean intensity values are calculated in this region across the time producing time-intensity plots.}
\label{dataset-DCE}
\end{figure*}

The functional kidney DCE-MRI datasets used in this study have been provided by collaborators at the Great Ormond Street Hospital using 1.5T Siemens Avanto scanner with 32-channel body phased-array coil. The datasets obtained are from ten healthy volunteers as shown in Figure~\ref{dataset-DCE}. The acquired DCE-MRI datasets cover the abdominal region, enclosing left and right kidneys and abdominal aorta. The anatomical images can be used to produce organ templates. The dataset consists of 120 MR images taken in sequence for $345$ seconds showing the central kidney slice i.e., the largest portion of kidney region. 


\subsection{Synthetic data}
In order to validate our experiments, and to demonstrate the capacity of our framework for functional separation,  we artificially generate synthetic data corresponding to broad DCE-MRI events in terms of simple mathematical functions. 

\begin{figure}[h]
\centering
\begin{tabular}{c}
\includegraphics[scale=0.4]{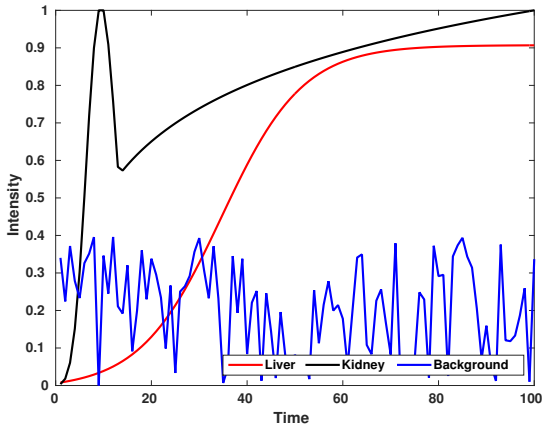} \\
\end{tabular}
\caption{Graph showing the time and intensity evolution of kidney, liver and background functions.}
\label{fulspec-1}
\end{figure}

\begin{figure}[h]
\centering
\begin{tabular}{c}
 \includegraphics[scale=0.5]{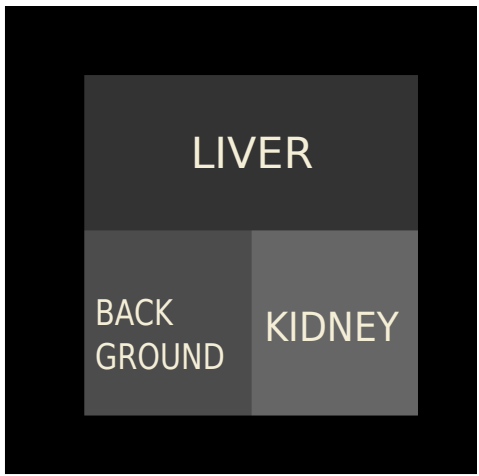}\\
\end{tabular}
\caption{Simulated image representing kidney, liver and background functions.}
\label{fulspec-2}
\end{figure}


A series of 100 images are produced representing the kidney, liver and background functions as labelled in Figure~\ref{fulspec-2} with temporal evolution shown in Figure~\ref{fulspec-1}(a). Kidney function is hence modelled as a linear combination of Poisson and log distributions; liver function is modelled as a sigmoid distribution and the background function is derived from a Gaussian noise  distribution.


\section{Experiments} 
\label{exp_res}
In this section we outline the experimental procedure as well as the corresponding results. 

\subsection{Evaluating the DMD framework on synthetically generated data.} We firstly investigate the performance of the proposed framework on the synthetically-generated data embodying  a coarse-grained  simulation of   DCE-MRI composite dynamics. This evaluation will  provide the proposed DMD framework with a ground truth  and establish the reproducibility and repeatability of our experimental results. Since DMD is completely data-driven, we conjecture that DMD will capture the kidney, liver as well as the background functions within its modes. 

\begin{figure*}
\centering
\begin{tabular}{ccc}
\includegraphics[scale=0.31]{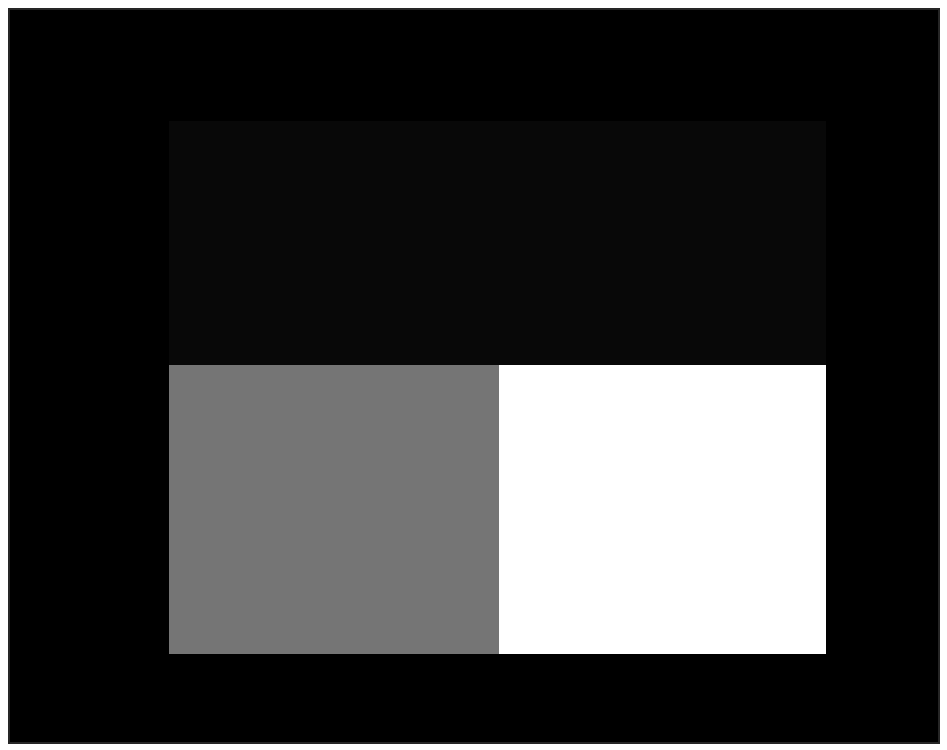}  & \includegraphics[scale=0.31]{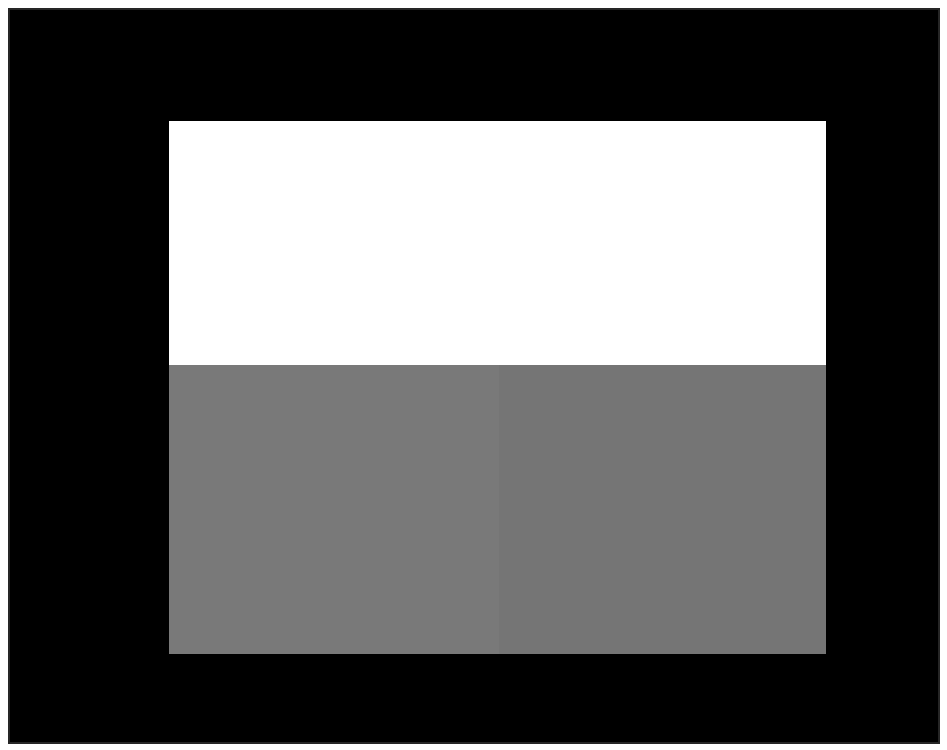}& \includegraphics[scale=0.31]{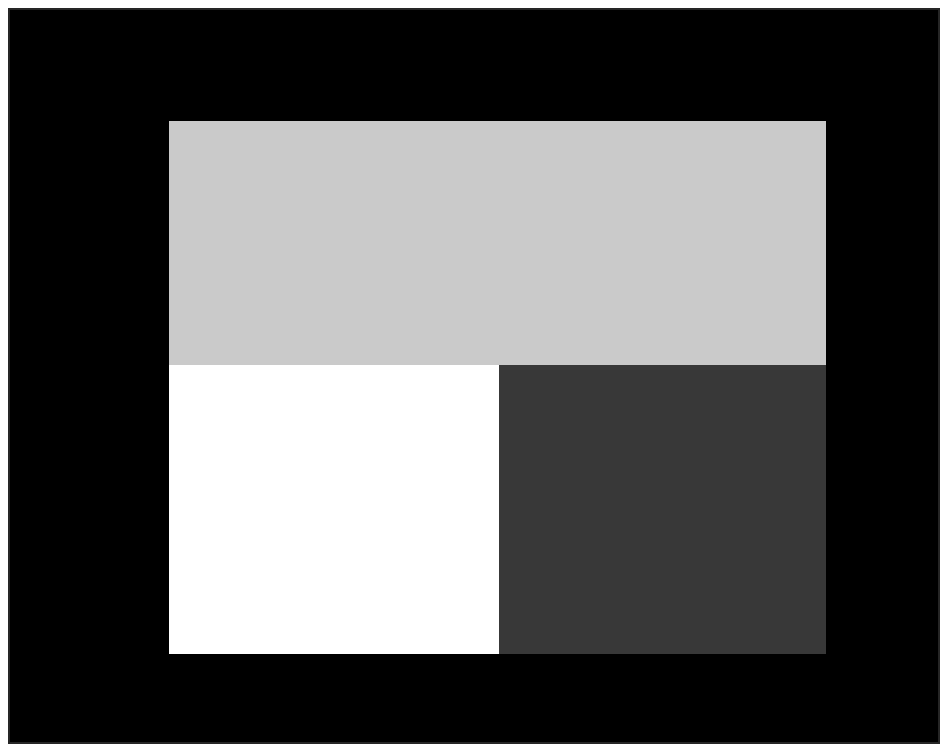}\\
DMD Mode-1 & DMD Mode-2 & DMD Mode-7\\
\includegraphics[scale=0.31]{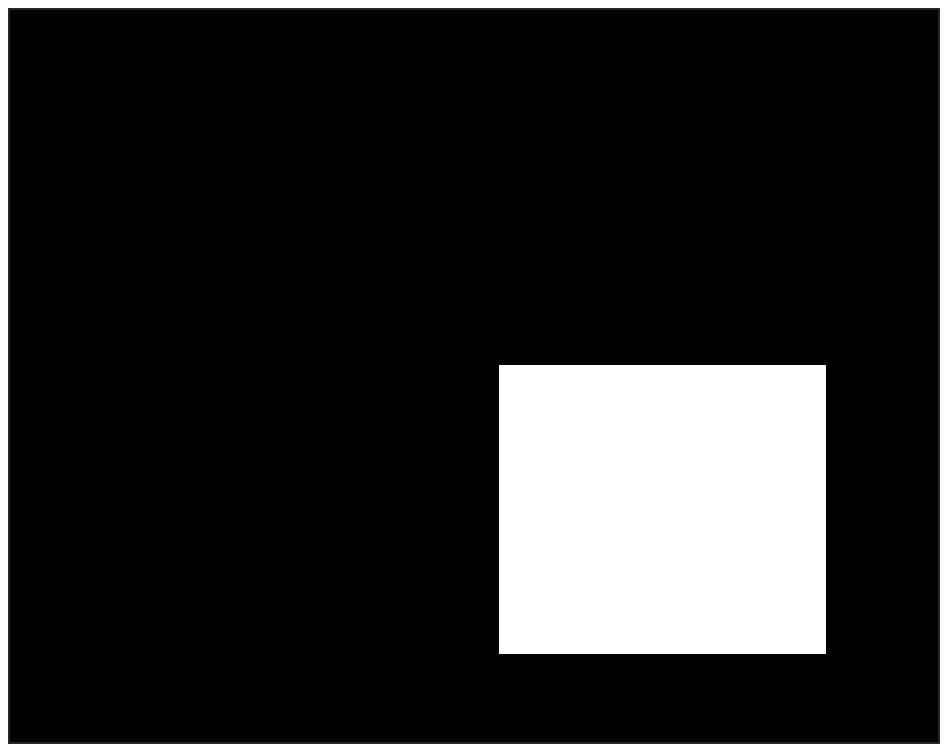}  & \includegraphics[scale=0.31]{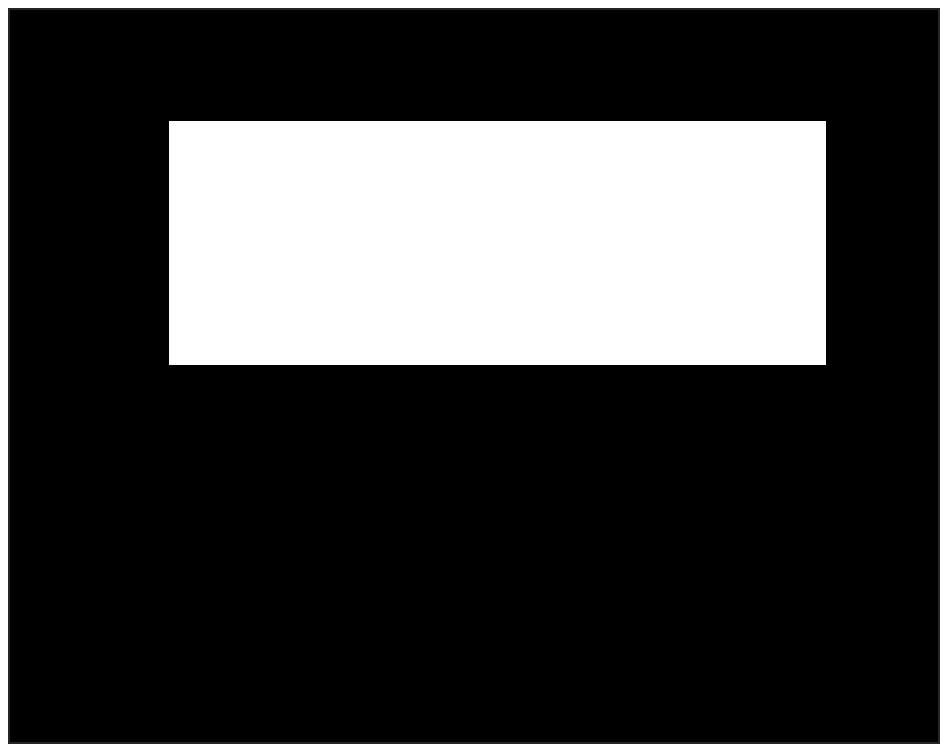}& \includegraphics[scale=0.31]{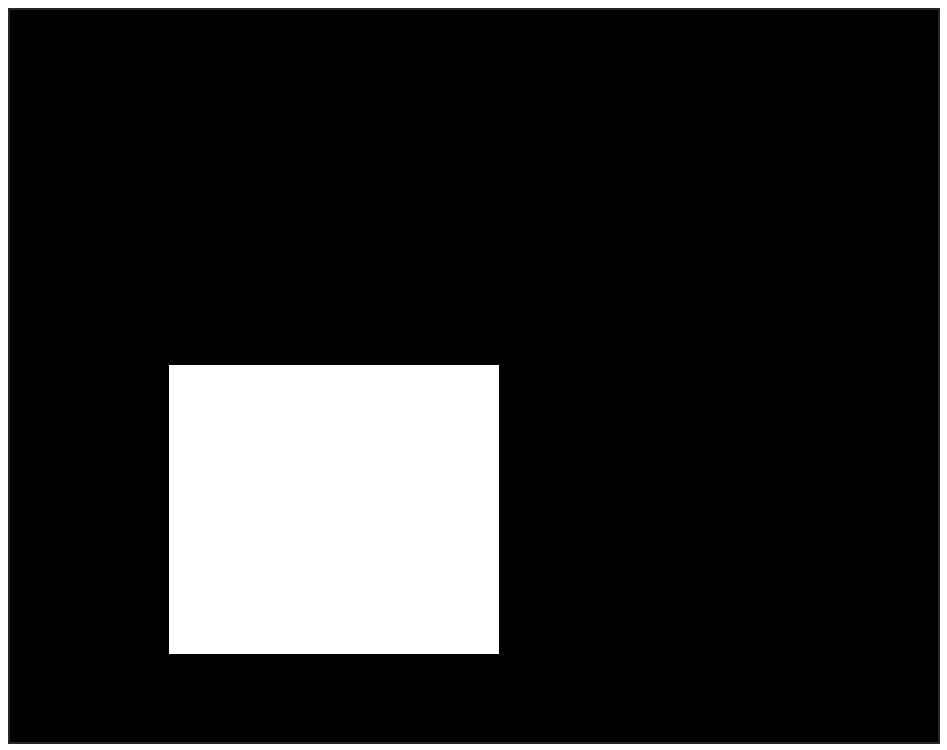}\\
Thresholded DMD Mode-1 & Thresholded DMD Mode-2 & Thresholded DMD Mode-7\\
\includegraphics[scale=0.3]{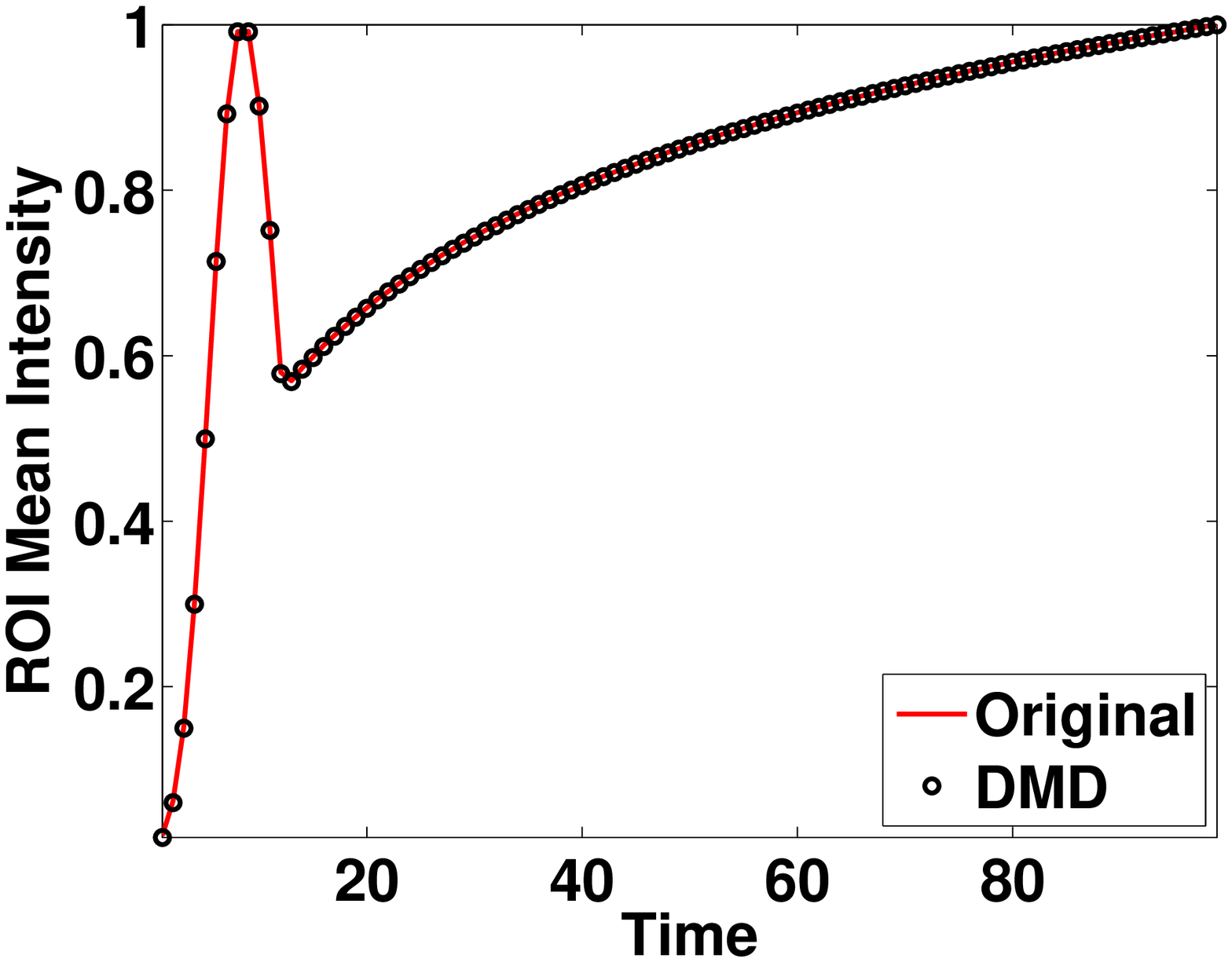}  & \includegraphics[scale=0.3]{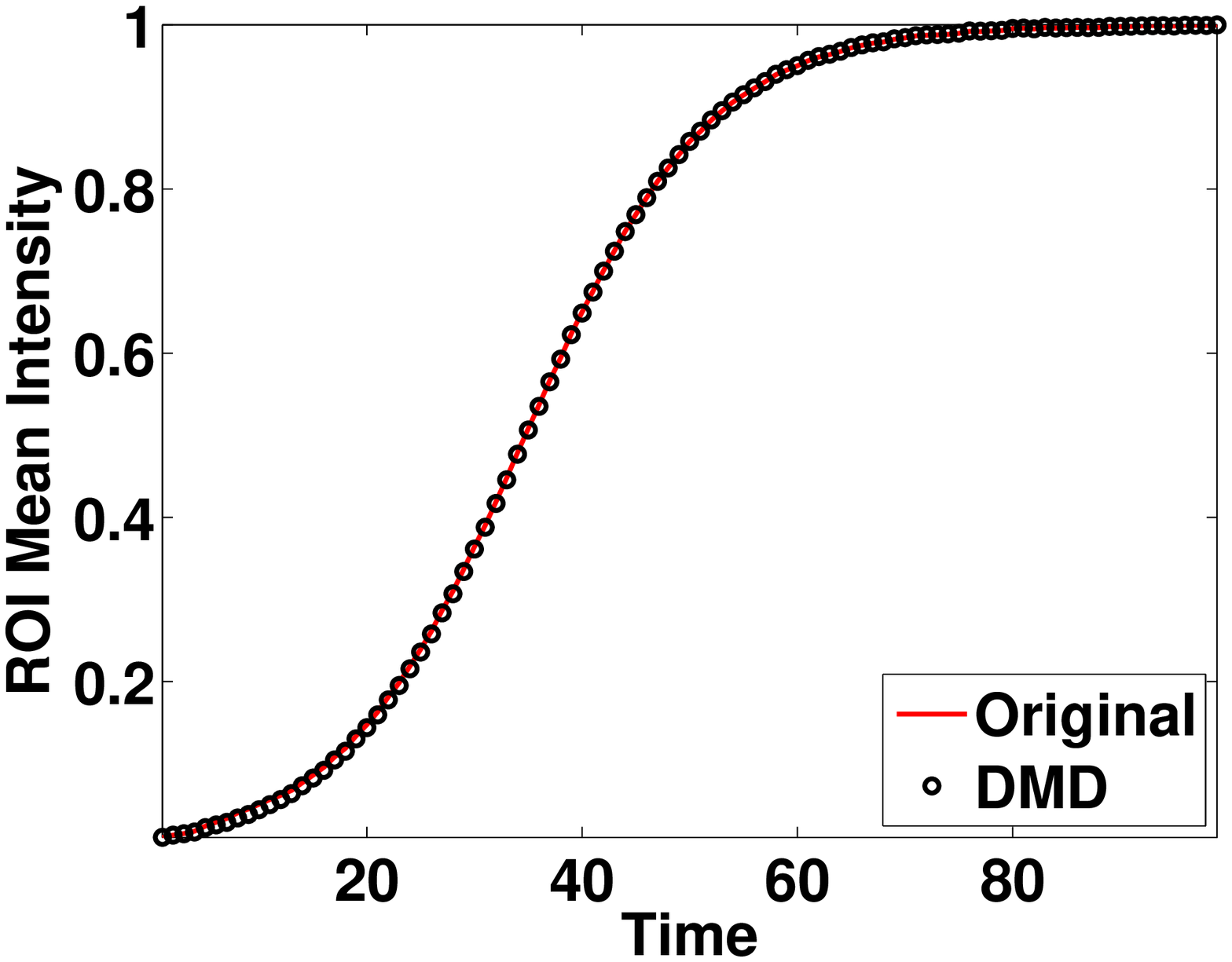} & \includegraphics[scale=0.3]{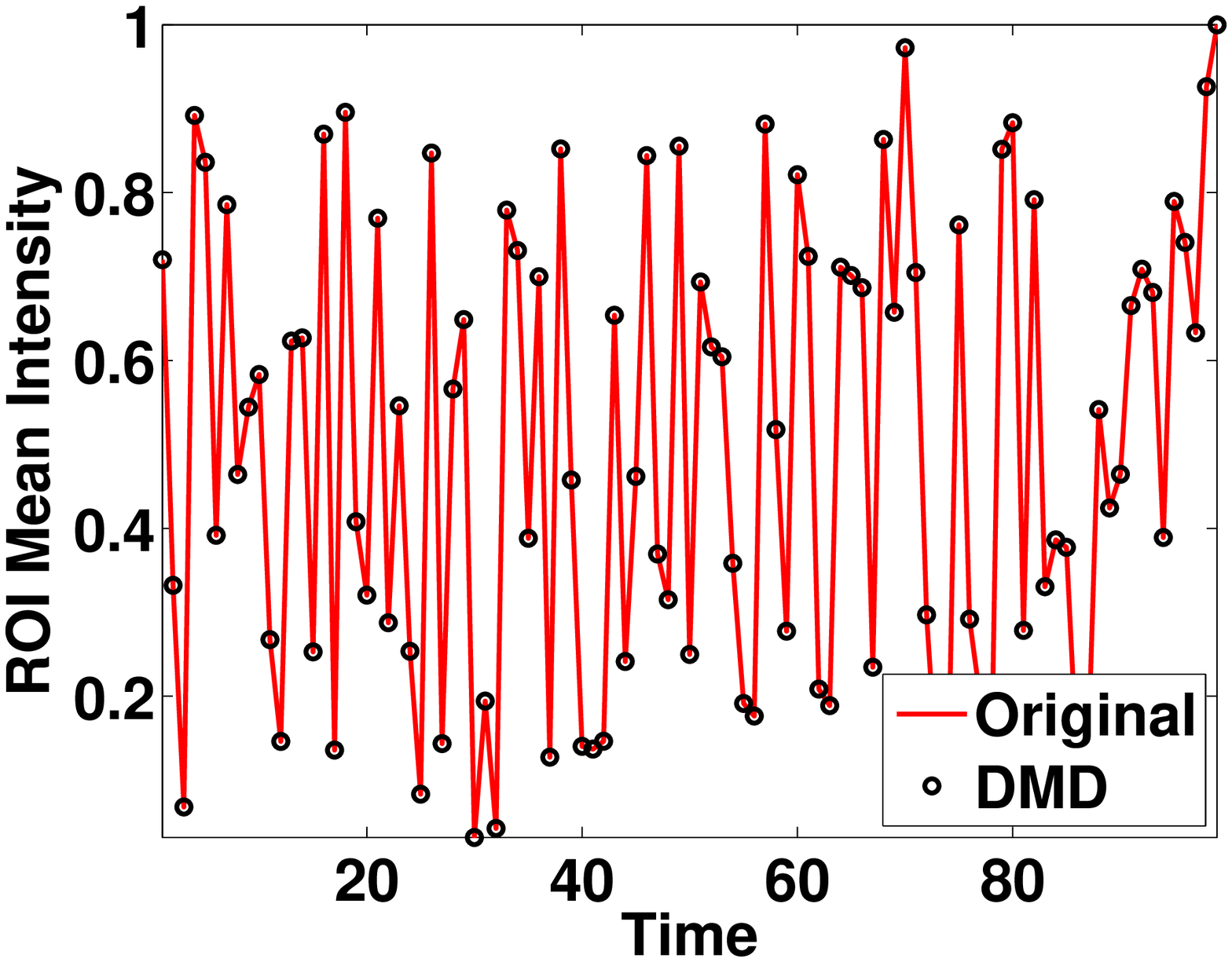}\\
(a) & (b) & (c)
\end{tabular}
\caption{Selection of DMD modes (top), thresholding (middle), extraction and quantification of (a) kidney, (b) liver and (c) background functions.}
\label{res:fig7}
\end{figure*}

Following the outlined  methodology, synthetic data consisting of a $100$ image sequence is given as an input to the modified DMD algorithm producing $99$ DMD modes (In theory, for a image sequence with $N$ images, we obtain $N-1$ DMD modes). Recall that, in principle, each  dynamic mode captures  one of the Key dynamic axes of the image sequence. Modes that show pre-dominating liver, kidney and background functions are manually selected, as shown in Figure~\ref{res:fig7} (top). These modes are then thresholded to obtain the segmented versions of respective functions as shown in Figure~\ref{res:fig7} (middle). As the DMD is completely data-driven, aside from  mode selection (in this case), we have thus demonstrated our conjecture that DMD can capture intact the distinct kidney, liver as well as the background functions from the synthetic data, irrespective of their intrinsically varied morphological  kinds. More generally, we have demonstrated the intrinsic capacity of DMD for {\it functional} separation as distinct from  ICA's {\it stochastic} component separation. We further characterise this separability by projecting the segmented versions of Dynamic modes 1, 2 and 7 for quantifying kidney, liver and background functions as shown in Figure~\ref{res:fig7}(bottom). The Jaccard Similarity Coefficient achieved for synthetic dataset is $1$ and Mean Square Error between the DMD result and the ground truth is $0.0000$.

\subsection{Implementation of the DMD framework on DCE-MRI data} Due to the injection of contrast agent, large scale intensity fluctuations can be observed inside the kidney region which also has minor impact on the liver region. We therefore wish to examine whether these features are indeed picked up by DMD. 

\begin{figure*}
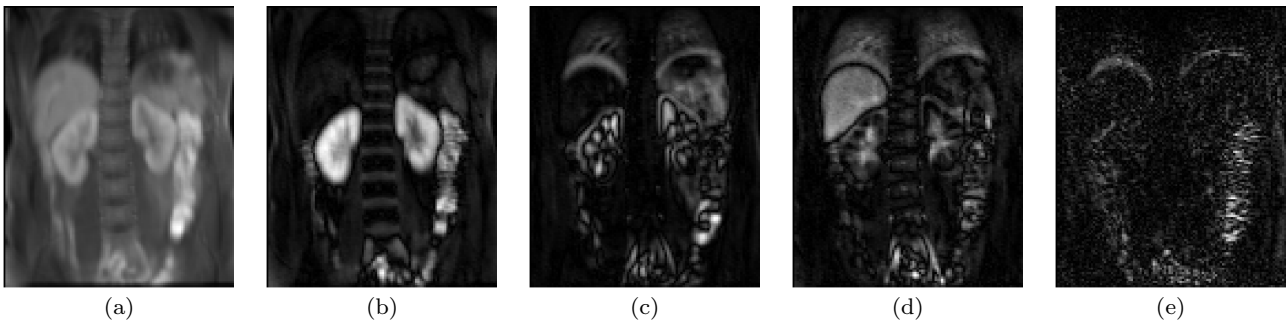

\centering
\begin{tabular}{ccccc}
\includegraphics[scale=0.30]{pictures/DCE/result_mode_1.eps} & \includegraphics[scale=0.30]{pictures/DCE/result_mode_2.eps} & \includegraphics[scale=0.30]{pictures/DCE/result_mode_4.eps} & \includegraphics[scale=0.30]{pictures/DCE/result_mode_5.eps} & \includegraphics[scale=0.30]{pictures/DCE/result_mode_83.eps}\\
(a) & (b) & (c) & (d) &(e)\\
\end{tabular}
\caption{DMD modes 1,2,4,5 and 83 from a health volunteer's data (dataset-1). (a) DMD mode-1 showing the background image captured from the dataset. (b) DMD mode-2 showing predominantly the kidney region. (c)  DMD mode-4 showing predominantly the spleen region. (d) DMD mode-5 showing predominantly the liver region. (e) DMD mode-83 showing the noise component from the dataset. }
\label{res:modes}
\end{figure*}

DCE-MRI image sequences from the healthy volunteers' data are given as  input to DMD algorithm. The output of  DMD is thus modally separated large and small scale fluctuations in voxel intensity. 
These modes are obtained by suppressing the stable information present within the DCE-MRI image sequences (with the exception of  mode 1); DMD is thus able to suppress the background information from the image sequence insofar as this is stationary. Renal perfusion (due to the injection of contrast agent) inside the kidney region exhibits large scale voxel intensity fluctuations, followed, in order of magnitude,  by liver and spleen region fluctuations respectively. DMD thus captures key kidney, liver and spleen regions in distinct modes as shown in Figure~\ref{res:modes}. Dynamic mode-1 captures the low-rank image of the DCE-MRI dataset-1 as shown in Figure~\ref{res:modes}(a) revealing the background function of the dataset. The kidney function is empirically observed to be captured in dynamic mode-2 (b) and the spleen and liver functions captured in dynamic mode-4 (c) and 5 (d) respectively. The lower order modes reveal the noise components inside the dataset-1 as seen in Figure~\ref{res:modes}(e). These results hence further provide visual evidence for our conjecture that DMD is capable of isolating key functional regions such as kidney and liver. These rankings however are not given intrinsically by DMD (as discussed previously); we now look at how this can be accomplished. 

\begin{figure*}
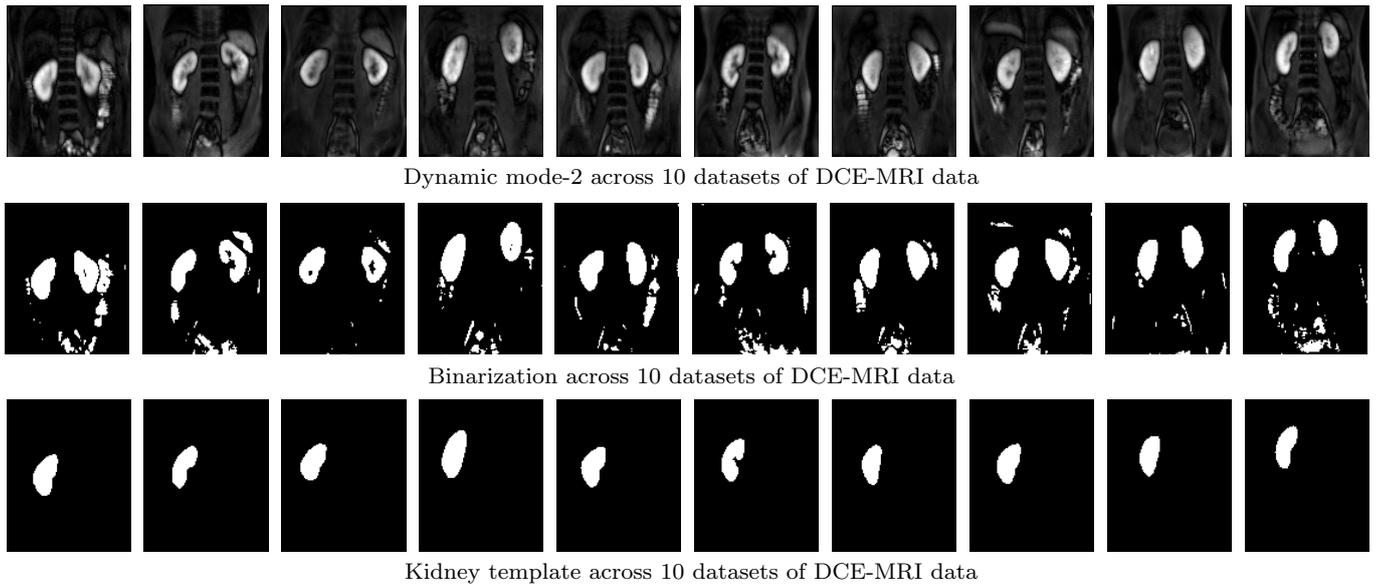

\centering
\begin{tabular}{c}
\includegraphics[scale=0.9]{pictures/DCE/dmd_mode_2.eps} \\
Dynamic mode-2 across 10 datasets of DCE-MRI data\\ 
\includegraphics[scale=0.9]{pictures/DCE/segmented_kidney.eps}\\ 
Binarization across 10 datasets of DCE-MRI data\\
\includegraphics[scale=0.9]{pictures/DCE/blobs.eps} \\ 
Kidney template across 10 datasets of DCE-MRI data\\
\end{tabular}
\caption{(Top) Image showing the kidney regions in Dynamic mode-2 across the 10 datasets used in this study. (Middle) Images showing the binarization effect on dynamic mode-2 across the datasets. (Bottom) Images showing the kidney template selected by the largest area of connected components. }
\label{res:fig2}
\end{figure*}

The main aim in this study is to segment the kidney region so as to enable quantification of  the kidney function. We thus select the second dynamic mode for this functional characterisation as this configuration is observed to produce kidney region  segmentation, consistent with the above argument,   across all of the datasets as shown in Figure~\ref{res:fig2} (top).  These dynamic mode-2s are first normalised in the range of $[0,1]$. Binarized images of dynamic mode-2s are then obtained by replacing all pixels in the image with intensity values greater than an adaptive threshold with the value 1 (white), and replacing all other pixels with the value 0 (black). These result in the binarized versions of  dynamic mode 2  shown in Figure~\ref{res:fig2} (middle). Kidney templates are then automatically selected as the largest area of  connected components of binarized dynamic mode-2 images across all the datasets as shown in Figure~\ref{res:fig2} (bottom). The connected components are selected by scanning the binarized image from top-to-bottom. All the pixels in a connected component are given a greedy label.  Thus, all the pixels in the first connected component are labelled as $1$ and those in the second as $2$ and so on. These components are then ordered by decreasing  area (i.e. the sum of all the pixels present in that particular component) and the label with the largest area is then selected as the kidney template. The produced kidney templates are then projected onto DCE-MRI images to obtain the kidney function (as shown in Figure~\ref{flowchart} from Section~\ref{method}) by calculating the mean intensity pixel values inside region specified by the template. The kidney functions across all the datasets obtained through the DMD framework are shown in Figure~\ref{res:fig4}.

\subsection{Evaluation of the DMD framework on DCE-MRI.} Evaluation of our framework is two-fold: 

\begin{itemize}
\item Evaluation of the performance of the segmented kidney template. 
\item Evaluation of the performance of the quantified  kidney function.
\end{itemize}

After obtaining the kidney ROIs (i.e the templates), the Jaccard Similarity Measure is  used to evaluate the performance of the proposed DMD framework with respect to the ground-truth data. The aim of this experiment is to benchmark the generalisation performance of our framework against expert annotations. 

For an unbiased evaluation, three different ground-truths $G_i$ $\forall \ \ i \in \{1,2,3\}$ are calculated based on the three different segmentation results obtained from the human experts. The ground-truth $G_1$ for the human expert-1 is calculated as the mutual agreement of the segmentation between expert-2 and expert-3. Similarly the ground-truth $G_2$ for expert-2 is calculated as the mutual agreement of segmentation between expert-1 and expert-3 and vice versa. We also have considered an additional  baseline given by the minimum bounding box region around the human annotated regions (to simulate 'blind' {\it a priori} segmentation). The DMD framework is evaluated as the mean of JSC values from all  three ground-truths. Table~\ref{tab_resdmd} shows the JSC values for the DMD framework against the three ground-truths  $G_i$ $\forall \ \ i \in \{1,2,3\}$. The last column of the table presents the mean of the JSC values obtained from the three ground-truths across all the 10 datasets.

\begin{table}[h]
\centering
\caption{Evaluation results for the DMD framework with respect to the three ground-truths $G_i$ $\forall i \in \{1,2,3\}$. }
\label{tab_resdmd}
\begin{tabular}{|l|l|l|l|l|}
\hline
        & \textbf{G\_1} & \textbf{G\_2} & \textbf{G\_3} & \textbf{avg(DMD)} \\ \hline
\textbf{Dataset-1}  & 0.8789        & 0.8672        & 0.8727        & 0.8729            \\ \hline
\textbf{Dataset-2}  & 0.8291        & 0.8525        & 0.8758        & 0.8525            \\ \hline
\textbf{Dataset-3}  & 0.8426        & 0.8628        & 0.8411        & 0.8488            \\ \hline
\textbf{Dataset-4}  & 0.9038        & 0.9133        & 0.9279        & 0.9150            \\ \hline
\textbf{Dataset-5}  & 0.8583        & 0.9027        & 0.8854        & 0.8821            \\ \hline
\textbf{Dataset-6}  & 0.8651        & 0.8418        & 0.8386        & 0.8485            \\ \hline
\textbf{Dataset-7}  & 0.8578        & 0.8239        & 0.8065        & 0.8294            \\ \hline
\textbf{Dataset-8}  & 0.8927        & 0.9075        & 0.8861        & 0.8954            \\ \hline
\textbf{Dataset-9}  & 0.8949        & 0.8702        & 0.8828        & 0.8826            \\ \hline
\textbf{Dataset-10} & 0.8356        & 0.8169        & 0.8799        & 0.8441            \\ \hline
\end{tabular}
\end{table}

The performance of the kidney segmentation achieved by the  DMD framework, the blind bounding-box region and three domain experts with respect to their ground-truths are reported in Table~\ref{tab:res1}. The JSC values achieved by expert-2 are generally high in  comparison to the other experts. The DMD framework has achieved higher JSC values  $0.91, 0.89 \ \& 0.88$ for the datasets 4 and 8 and 9. The Overall average JSC values for the experts are $0.85, 0.87 \ \& 0.87$ and for the DMD framework is around $0.86$ while for minimum bounding-box region the JSC  is below $0.04$

\begin{table}[h]
\centering
\caption{Jaccard similarity coefficients for kidney segmentation achieved by the DMD framework, blind bounding-box region and three domain experts with respect to the ground-truth}
\label{tab:res1}
\begin{tabular}{|l|l|l|l|l|l|}
\hline
\textbf{}           & \textbf{DMD} & \textbf{E-1} & \textbf{E-2} & \textbf{E-3} & \textbf{Bbox} \\ \hline
\textbf{D-1}  & 0.8729            & 0.8642            & 0.8994   & 0.8974            & 0.5891      \\ \hline
\textbf{D-2}  & 0.8525            & 0.8718            & 0.8925   & 0.8906            & 0.4563      \\ \hline
\textbf{D-3}  & 0.8488            & 0.8642            & 0.8781   & 0.8719            & 0.4298      \\ \hline
\textbf{D-4}  & 0.9150   & 0.8880            & 0.8894            & 0.8894            & 0.4073      \\ \hline
\textbf{D-5}  & 0.8821            & 0.8668            & 0.9064   & 0.8934            & 0.3162      \\ \hline
\textbf{D-6}  & 0.8485            & 0.8589            & 0.8679   & 0.8536            & 0.3016      \\ \hline
\textbf{D-7}  & 0.8294            & 0.8548            & 0.8833   & 0.8529            & 0.2692      \\ \hline
\textbf{D-8}  & 0.8954   & 0.8155            & 0.8726            & 0.8708            & 0.2984      \\ \hline
\textbf{D-9}  & 0.8826   & 0.8221            & 0.8711            & 0.8391            & 0.2624      \\ \hline
\textbf{D-10} & 0.8441            & 0.8426            & 0.8292            & 0.8792   & 0.2455      \\ \hline
\textbf{avg}    & 0.8671            & 0.8548            & 0.8789   & 0.8738            & 0.3575      \\ \hline
\end{tabular}
\end{table}

\begin{figure*}
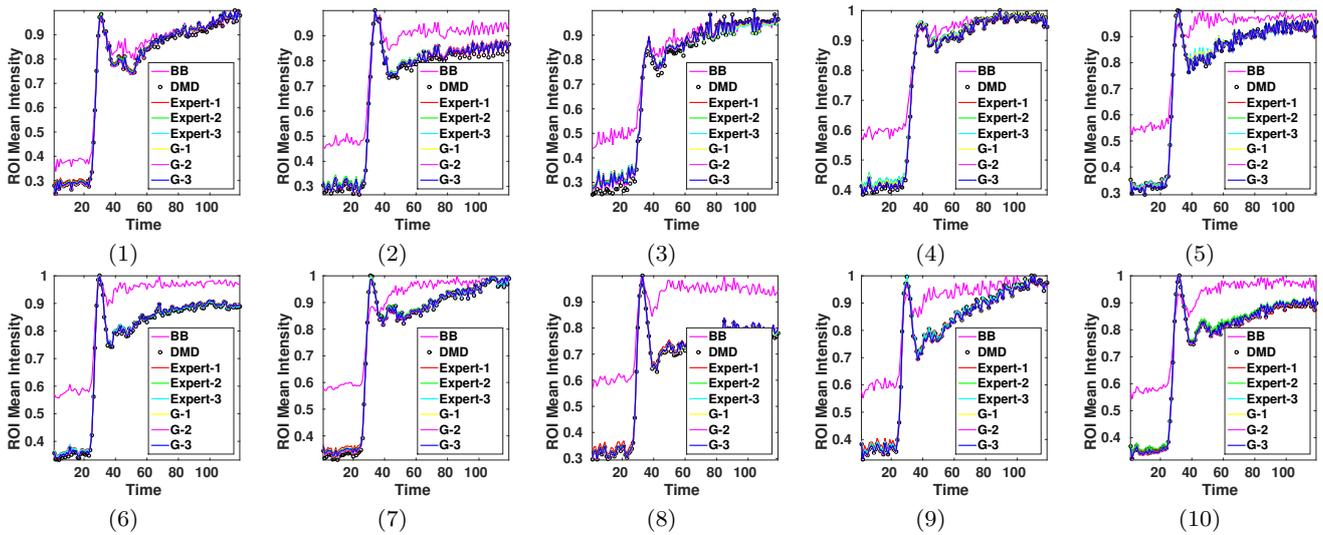

\centering
\begin{tabular}{ccccc}
\includegraphics[scale=0.21]{pictures/DCE/dataset-1.eps} & \includegraphics[scale=0.21]{pictures/DCE/dataset-2.eps} & \includegraphics[scale=0.21]{pictures/DCE/dataset-3.eps} & \includegraphics[scale=0.21]{pictures/DCE/dataset-4.eps} & \includegraphics[scale=0.21]{pictures/DCE/dataset-5.eps} \\
(1) & (2) & (3) & (4) & (5) \\
\includegraphics[scale=0.21]{pictures/DCE/dataset-6.eps} & \includegraphics[scale=0.21]{pictures/DCE/dataset-7.eps} & \includegraphics[scale=0.21]{pictures/DCE/dataset-8.eps} & \includegraphics[scale=0.21]{pictures/DCE/dataset-9.eps} & \includegraphics[scale=0.21]{pictures/DCE/dataset-10.eps} \\
(6) & (7) & (8) & (9) & (10)\\
\end{tabular}
\caption{Kidney function produced by blind bounding-box region, DMD framework, three domain experts and ground-truths for datasets 1-10.}
\label{res:fig4}
\end{figure*}

Figure~\ref{res:fig4} shows the kidney functions produced by the  DMD framework, ground-truth ($G_i$ $\forall \ \ i \in \{1,2,3\}$), three domain experts as well as kidney function produced by blind bounding-box over the kidney region. The results in these figures show that kidney function quantified by DMD is closely aligned with the experts annotation. 

Evaluation for the kidney functions are calculated based on the Mean Square Error (MSE) criteria as shown in Table~\ref{mse}. 

\begin{table}[h]
\centering
\caption{Mean Square Error (MSE) based evaluation of the kidney functions for DMD, human experts and the bounding box.}
\label{mse}
\begin{tabular}{|c|p{1cm}|c|c|c|p{1cm}|}
\hline
                    & \textbf{ avg (DMD)} & \textbf{E-1} & \textbf{E-2} & \textbf{E-3} & \textbf{ avg (Bbox)} \\ \hline
\textbf{D-1}  & 0.0001       & 0.0001            & 0.0001            & 0.0000            & 0.0029        \\ \hline
\textbf{D-2}  & 0.0001       & 0.0001            & 0.0001            & 0.0000            & 0.0126        \\ \hline
\textbf{D-3}  & 0.0002       & 0.0001            & 0.0001            & 0.0000            & 0.0082        \\ \hline
\textbf{D-4}  & 0.0000       & 0.0000            & 0.0001            & 0.0001            & 0.0085        \\ \hline
\textbf{D-5}  & 0.0001       & 0.0000            & 0.0000            & 0.0000            & 0.0166        \\ \hline
\textbf{D-6}  & 0.0000       & 0.0000            & 0.0000            & 0.0000            & 0.0188        \\ \hline
\textbf{D-7}  & 0.0001       & 0.0001            & 0.0001            & 0.0000            & 0.0162        \\ \hline
\textbf{D-8}  & 0.0001       & 0.0002            & 0.0000            & 0.0000            & 0.0431        \\ \hline
\textbf{D-9}  & 0.0000       & 0.0002            & 0.0001            & 0.0000            & 0.0176        \\ \hline
\textbf{D-10} & 0.0000       & 0.0000            & 0.0002            & 0.0001            & 0.0213        \\ \hline
\end{tabular}
\end{table}

drvdk144@gmail.com

\section{Discussions \& Conclusions}
\label{conc}

This study aims to demonstrate the significance of the proposed DMD framework   as a viable functional segmentation algorithm when
coupled with simple thresholding-binarization and selection of the largest area of connected pixels to effectively quantify kidney function in DCE-MRI data. We applied the DMD framework to10 sets of DCE-MRI data collected from healthy volunteers. We also applied the DMD framework to  synthetically generated data mimicking the DCE-MRI data. The results demonstrate that the proposed framework is extremely promising in obtaining functional segmentation in general and thus to quantifying segmented functions, in particular  kidney functionality.

DMD can be demonstrated to extract local variations as a low rank representation within an image sequence, as well as capturing dominating regions of causally-connected intensity fluctuations. In our context, perfusion inside the kidney region is the most dominating region of intensity fluctuations due to the injection of contrast agent.  DMD was thus able to naturally capture the kidney region as mode-2. Using a simple thresholding technique and selecting the largest area with connected pixels then automatically generates the proposed kidney region. The results for the proposed  DMD framework when compared with expert annotations and ground-truth clearly shows the strength of the framework suggesting that manual selection of kidney region is no longer needed and    that the entire process can be automated.

\begin{acknowledgements}
The funding for this work has been provided by the Department of Computer Science and the Centre for Vision, Speech and Signal Processing (CVSSP) - University of Surrey. `I.G' would like to express gratitude towards Kidney Research UK for funding the DCE-MRI data acquisition as part of a reproducibility study. `S.T' and `N.P' have benefited  from  the  Medical  Research  Council  (MRC) funded project ``Modelling the Progression of Chronic Kidney Disease'' under the grant number R/M023281/1. The details of the project are available at \url{www.modellingCKD.org}. `D.W' acknowledges the financial support from the Horizon 2020 European Research project ``DREAMS4CARS" (\#731593).
\end{acknowledgements}

\bibliographystyle{spmpsci}      
\bibliography{myPaper,DMD_ref,myPaper_ckd,santosh_references,confirmation,movement_correction,SBI}
%
%

\end{document}